\newif\ifarxiv
\arxivtrue 

\ifarxiv
    \documentclass{article}
\fi

\usepackage{lineno,hyperref,float}
\usepackage{pgfplots, pgfplotstable}
\usepackage{multirow}
\usepackage{multicol}
\usepackage{xcolor}
\usepackage{ifthen}
\usepackage{datetime}
\date{February, 2022}
\ifarxiv
    \usepackage[affil-it]{authblk}
\fi

\newcolumntype{C}[1]{>{\PreserveBackslash\centering}p{#1}}

\modulolinenumbers[5]

\ifarxiv
\fi

\begin{document}

\title{Isoform Function Prediction Using a Deep Neural Network}

\ifarxiv
    \author{Sara Ghazanfari$^{1,2}$%
	    \thanks{e-mail: \texttt{sghazanfari@ce.sharif.edu}}}
    \author{Ali Rasteh$^{3,}$}
    \author{Seyed Abolfazl Motahari$^{1,}$}
    \author{Mahdieh Soleymani Baghshah$^{2,}$}

    \affil{\normalsize $^1$Bioinformatics Research Laboratory, Computer Engineering Department, Sharif University of Technology, Tehran, Iran}
    \affil{\normalsize $^2$Machine Learning Laboratory, Computer Engineering Department, Sharif University of Technology, Tehran, Iran}
    \affil{\normalsize $^3$Artificial Creatures Laboratory, Electrical Engineering Department, Sharif University of Technology, Tehran, Iran}
    
    \maketitle
\fi

\begin{abstract}
\emph{Isoforms} are mRNAs produced from the same gene site in the phenomenon called Alternative Splicing. Studies have shown that more than 95\% of human multi-exon genes have undergone alternative splicing. Although there are few changes in mRNA sequence, They may have a systematic effect on cell function and regulation. It is widely reported that isoforms of a gene have distinct or even contrasting functions. Most studies have shown that alternative splicing plays a significant role in human health and disease. Despite the wide range of gene function studies, there is little information about isoforms' functionalities. Recently, some computational methods based on Multiple Instance Learning have been proposed to predict isoform function using gene function and gene expression profile. However, their performance is not desirable due to the lack of labeled training data. In addition, probabilistic models such as Conditional Random Field (CRF) have been used to model the relation between isoforms. This project uses all the data and valuable information such as isoform sequences, expression profiles, and gene ontology graphs and proposes a comprehensive model based on Deep Neural Networks. The UniProt Gene Ontology (GO) database is used as a standard reference for gene functions. The NCBI RefSeq database is used for extracting gene and isoform sequences, and the NCBI SRA database is used for expression profile data. Metrics such as Receiver Operating Characteristic Area Under the Curve (ROC AUC) and Precision-Recall Under the Curve (PR AUC) are used to measure the prediction accuracy.
\end{abstract}


\section{Introduction}
\label{sec:Introduction}
\begin{figure}[]
\centering
\includegraphics[%
width=1.0\textwidth]{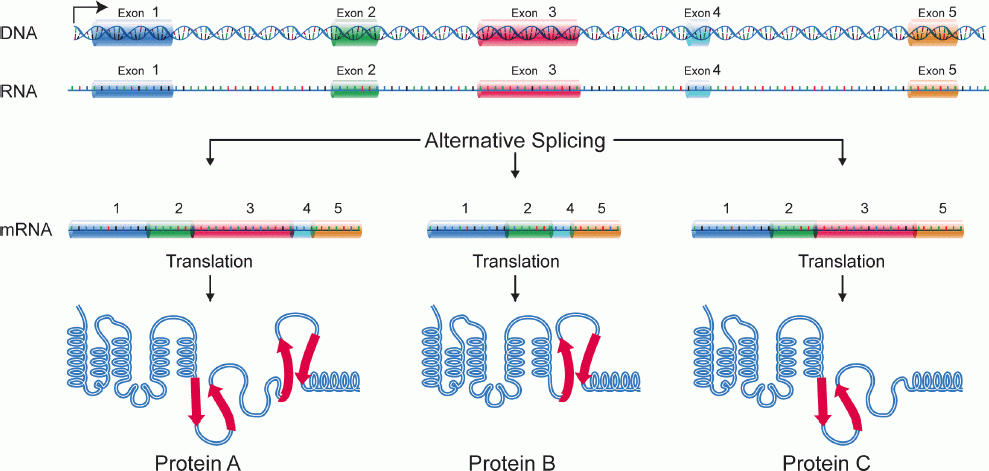}
\caption{\label{fig:dna_alternative_splicing}{Alternative Splicing\cite{ellis2012tissue}. Only gene exons are transcribed to mRNAs in the transcription process, and any combination of transcribed exons can produce an mRNA. This event is called Alternative Splicing.}}
\end{figure}

DNA molecules encode the biological information that determines instructions for development and functioning inside the cells of all known organisms. Each function is encoded on a specific part of DNA called a gene. In other words, genes are substrings of the DNA sequence. Each gene is made up of two regions, Exons and Introns. According to The Central Dogma of Molecular Biology, DNA transcribes to RNA, translating to protein. Although introns do not transcribe to RNA, exons participate in the transcription process, and any combination of them can produce RNA. This process which leads to the production of multiple RNAs and proteins from the same gene, is called Alternative Splicing. The different RNA that is produced from the same gene is called isoform. Alternative Splicing is a source of diversity in organisms’ functions. For instance, it is estimated that humans have between 20,000 and 25,000 genes. Although this number is smaller than the number of functions, the processes like Alternative Splicing in the central dogma make the diversity of functions possible. A schematic illustration of Alternative Splicing, as well as central dogma, is given in Figure \ref{fig:dna_alternative_splicing}.

In recent years, research in genomic functions has transferred from the gene level to the isoform level. The research results have shown that isoforms of a gene may have different or even contrasting functions. For example, the BCL2L1 gene produces two isoforms with contradicting functions for the apoptosis of tumor cells. BCL-XS and BCL-XL isoforms are pro-apoptosis and anti-apoptosis, respectively\cite{revil2007protein}. Similar to this example is the CASP3 gene which produces two isoforms with contradicting functions\cite{vegran2006overexpression}. There are other examples of isoforms with dissimilar functions in the literature\cite{himeji2002characterization,melamud2009stochastic,oberwinkler2005alternative,pickrell2010understanding}. Therefore, it is essential to investigate the genomic functions with smaller granularity and more accuracy in the isoform level.

Recently, several methods have been proposed for isoform function prediction, including iMILP\cite{li2014high}, WLRM\cite{luo2017functional}, DeepIsoFun\cite{shaw2019deepisofun}, and DIFFUSE\cite{chen2019diffuse}. All of them overcome the lack of labeled data by distributing the functional annotation of a gene to all of its isoforms using techniques such as multiple instance learning (MIL) and domain adaptation. Expression profiles are a common source of data in all proposed methods. However, the isoform sequence and conserved domains were first used by DIFFUSE, which led to a significant increase in the performance of the proposed model.

In this research, the goal is to predict functions of isoforms using their biological features and data, including isoform sequence, expression profile, and conserved domains. Due to the diversity and the large number of isoforms, exploring their functions through lab-based methods is impossible. As a result, scalable computational methods with accurate results for predicting isoform functions are demanding. On the other hand, training models require labeled data. Although there is no labeled data at the isoform level, comprehensive training data has been collected at the gene level. In the proposed method, we transfer the information in the gene level to isoform level using a Semi-supervised learning approach to predict the functions of isoforms.

\section{Materials and methods}
\label{sec:Materials and methods}
\subsection{Dataset}
\label{sec:Dataset}
\begin{figure}[]
\centering
\includegraphics[%
width=0.6\textwidth]{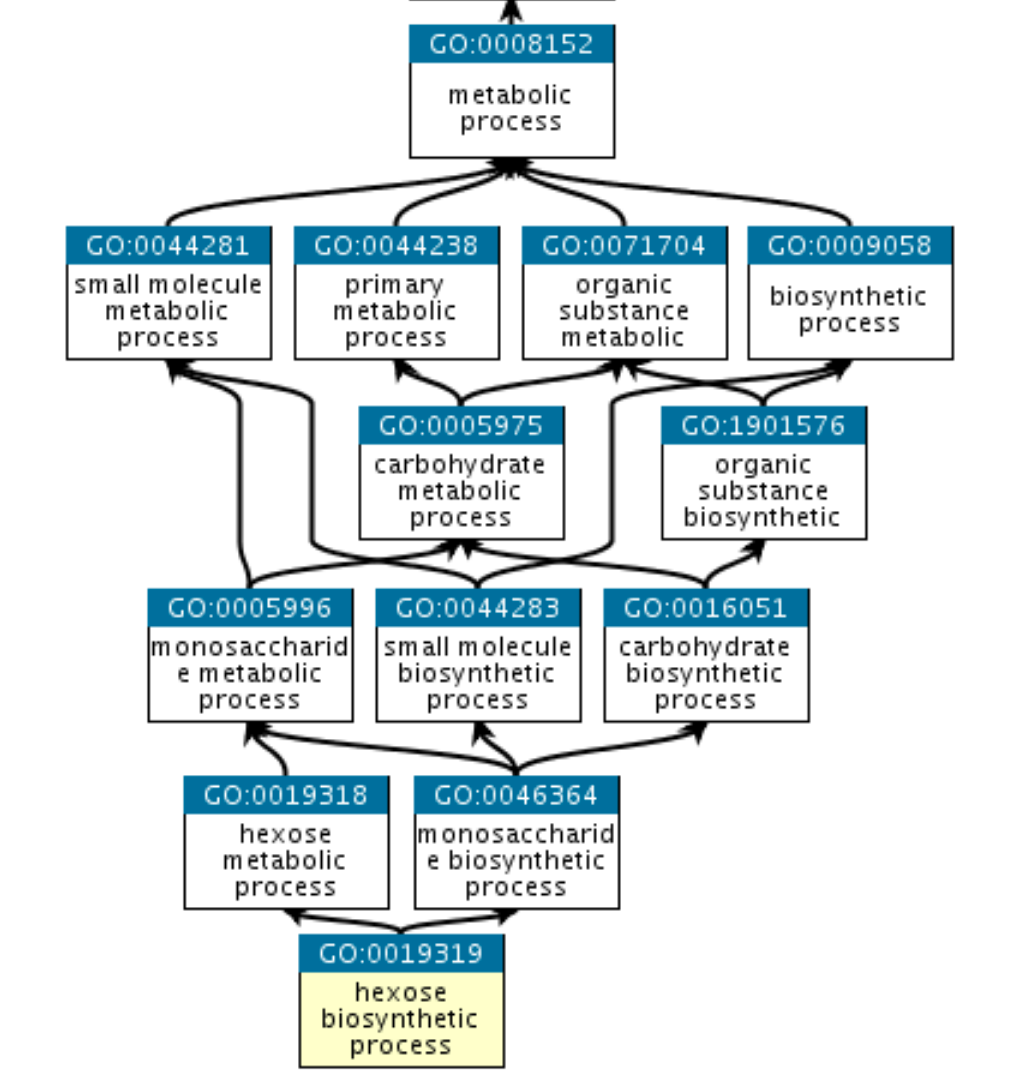}
\caption{\label{fig:go}{GO Terms' Graph. The GO terms have a graph structure; nodes represent the functions, and edges represent the relations. The child GO terms are specific sub-functions from the parent GO term.}}
\end{figure}

The input data for the model consists of isoform sequence, conserved protein domains, and expression profiles. These data have been collected and contain 39,375 isoforms from 19,303 genes. The expression profile data consists of 334 studies and 1,735 experiments. The output data, which is the gene functions come from a standard database named UniProt Gene Ontology. This database has a graph-based structure, which is shown in Figure \ref{fig:go}. The graph's nodes are functions (GO), and edges determine how the function divides into sub-function. The graph consists of three main branches, Biological Processes (BP), Molecular Functions (MF), and Cellular Components (CC). The functions become detailed when traversing from the root to the leaf of the graph. Genes are mapped to the GOs corresponding to their function. The total number of GOs in this research is 4,184 associated with selected genes. 

\subsection{Methods}
\label{sec:Methods}
\subsubsection{Protein sequence embedding}
\label{sec:Protein Sequence Embedding}
We use protein sequences in two ways, raw protein sequence and conserved domains, which should be embedded into numerical space. Embeddings are components in the deep neural network model that can be pre-trained (static) or trainable (dynamic). Dynamic embeddings are trained in the process of training the whole model. Therefore, the trained embeddings are problem-specific and can lead to better results in prediction. However, It is noteworthy that embedding components increases the model parameters, and in the case of high-dimensional embeddings and lack of sufficient data can lead to severe overfitting. In the case of finding embeddings for protein sequence, the number of parameters of the dynamic embedding is outstanding. The protein sequence is divided into overlapping subsequences with three characters in the conventional dynamic embedding for protein. As the number of amino acids is 20, the total subsequences are $20\times20\times20$, equal to 8,000. The average length of protein sequences in our samples is 3,000. As a result, the parameters number is 24 million ($8,000\times3,000$), which is a significant. Although dynamic embedding has been used in the literature, it is inappropriate for our problem. Our choice for embedding protein sequences is to use a pre-trained embedding named Protvec\cite{asgari10protvec}. On the other hand, conserved domains are limited in number (16,000), and it is better to train a dynamic embedding for it. 

\subsubsection{Proposed model}
\label{sec:Proposed Model}
The proposed method predicts isoform functions using features extracted from isoform sequence, conserved domains, and expression profile. The isoform sequence is fed to the Protvec component, producing an embedding. Afterward, a CNN component is used to extract sequence features and a dense layer to output the compact feature vector. On the other side, the conserved domains are injected into a dynamic embedding component, and sequence data is converted to numerical data. As the sequence of conserved domains contains information, an LSTM module is used to capture the information in the order of conserved domains. Another source of information comes from expression profiles. The expression profile is fed to a normalization component to remove the noise from the experiments and then fed to the dense layer to extract the feature vector. The feature vectors from the initial data are concatenated to a single feature vector. After passing the aggregated feature vector from a dense layer, the isoform level scores for GO terms are generated. However, as mentioned earlier, the labeled data is at the gene level. A customized max-pooling layer transforms the isoform level score to the gene level score. This component is an innovation from our work based on the formulation in the Multiple-Instance Learning approach.

In the MIL, the instances do not have labels. The instances are mapped to their related bag, and the label of bags is known. A bag has a positive label if at least one of the instances mapped to the bag has a positive label. In other words, if a bag has a negative label, it means none of the instances have a positive label in the bag. In this formulation, the maximum score of the instances leads to the score of the bag. This approach is well-matched with our current problem. A gene has a function if at least one of the isoforms has the function. As a result, the customized max-pool is designed to get the maximum score of the gene’s isoforms and assign it to the specific gene. By applying this component, the scores are transferred from the isoform level to the gene level, and the training process can be performed. 

\begin{figure}[]
\centering
\includegraphics[%
width=1.0\textwidth]{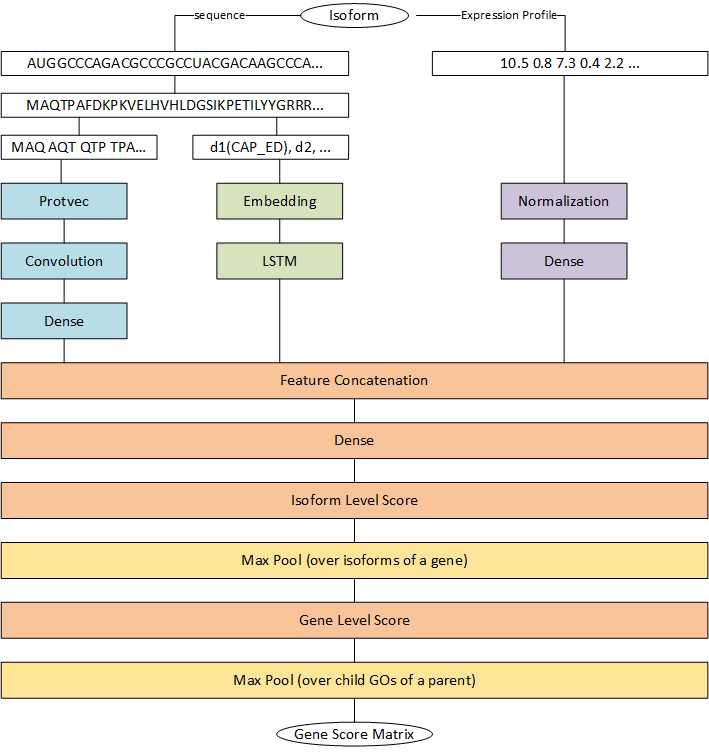}
\caption{\label{fig:sara_model}{Proposed model's architecture. Isoforms sequence and expression profile are given to the network as input. On the left side, the isoforms sequence is translated to the protein sequence and embedded using the Protvec component. The other part of the feature vector is generated by passing the embedded vector through the CNN and dense layers. The conserved domains are extracted from the protein sequence, and after applying the LSTM component, the second part of the feature vector is extracted. On the right side, the expression profile is normalized and fed to a few dense layers, extracting the feature vector's final part. Finally, three extracted feature vectors are concatenated and fed to dense layers to produce the isoform level score. Afterward, the isoform level scores are transferred to the gene level by applying the customized max-pooling layer over the isoforms of a gene. The final gene-level scores are produced by considering the GO terms' graph.}}
\end{figure}

In the last component, inspired by DeepGO\cite{kulmanov2018deepgo}, the structure of GO terms is used to increase the model's performance. As mentioned earlier, the GO terms have a graphical structure. By traversing from root to leaves, the GO terms get detailed. As a result, if a gene has a function associated with the child GO term, it has the function of the parent GO term. Another customized max-pooling layer is added to include this information in the model, which gets the maximum score between the child and parent GO term and assigns it to the parent GO term score. See Figure \ref{fig:sara_model} for the detailed NN architecture, and Figure \ref{tab:Pseudocode} for algorithm pseudo-code.

Despite the previous methods that train a separate model for each GO term, our method trains a model for each GO term in one of the three main branches, totally three models. The pros and cons of this approach are mentioned in the next section.

\begin{center}
\begin{figure}
\begin{tabular}{|l|}
\hline
Algorithm Pseudo-code  \\
\hline
\textbf{Inputs:} isoform sequence (s), conserved domain (d) and expression profile (e)\\
\hline
\textbf{Output:} isoform functions, $\hat{y}$\\
\hline
1: Compute score matrix in isoform level, $\hat{y}=DNN(s, d, e)$\\
2: for each gene i do \\
3:\hspace{4pt} for each gene function j do \\ 
4:\hspace{8pt} Compute function matrix in gene level, $\hat{y}[i][j] = \max_{k \in isoforms_{i}}{\hat{y}[k][j]}$ \\
5:\hspace{4pt} end for \\
6: end for \\
7: for each go i do \\
8: \hspace{4pt} Consider the go hierarchy by $\hat{y}[:][i] = max(\hat{y}[:][i], \hat{y}[:][child_i])$ \\
9: end for \\
10: Use SGD to update w\\
\hline
\end{tabular}
\caption{\label{tab:Pseudocode}Algorithm Pseudo-code for the proposed method.}
\end{figure}
\end{center}

\subsubsection{Training method}
\label{sec:Training Method}
Isoform Function Prediction is a Classification problem. The instances are isoforms, and the classes are functions. As the isoforms can have more than one function, the problem is a multi-label classification. One of the challenges in such a problem is an unbalanced dataset when negative samples for a class are much more than positive ones. In our problem, the number of classes is more than 4000, and positive samples for some classes are less than 10. The previous methods tried to overcome this challenge by training a separate model for every class. Therefore they can up-sample the positive and down-sample the negative samples to prepare a balanced dataset. The mentioned solution has some disadvantages:

\begin{itemize}
    \item Training more than 4,000 models, tuning the hyper-parameters, and saving the trained models require much time and computational resources.

    \item As the functions are related to each other, and it is possible to assign a group of functions to genes, separating models for each gene leads to information loss and performance decay.

    \item Due to data imbalance and insufficiency, the demand for sharing data and knowledge gets critical.

    \item As the network has many parameters, training each function in a separate network increases the overfitting possibility and decreases generalization.
\end{itemize}

To train all functions in the same model and gain the benefits that are stated above, we first need to overcome the data imbalance problem. To overcome this problem, the training technique which is used in multi-task learning is applied. In each training iteration, first classes and then instances associated with the chosen class are sampled in a balanced way. As a result, the gradient for chosen classes is updated in each iteration. 

The rest of this section will explain the loss function and how it was modified to make the training possible. Generally, the loss function of the Multi-Task Learning solution is composed of the sum of loss functions of each of the tasks shown in equation \ref{eq:1}.

\begin{center}
\begin{equation}
L_{tot}(X, Y_{1:K}) = \sum_{i=1}^{K}\lambda_{i}L_{i}(X,Y_i)
\label{eq:1}
\end{equation}   
\end{center}

Where $\lambda_{i}$ is the weight of task i. One of the challenges in the MTL is finding the balance between tasks. The balance can be achieved by determining the optimum value for the weights. We used the Dynamic Weight Average\cite{liu2019end} method to find the best weights, which are calculated by considering the loss gradient for each task. Despite other methods that require the gradients inside the network for their calculations, DWA only needs the loss value as shown in equation \ref{eq:2}.

\begin{center}
\begin{equation}
\lambda_k(t) := \frac{K exp(w_k(t-1)/T)}{\sum_i exp(w_i(t-1)/T)}, w_k(t-1) = \frac{L_k(t-1)}{L_k(t-2)}
\label{eq:2}
\end{equation}   
\end{center}

The hyper-parameter T is used to determine the smoothness between tasks’ weights. Large T leads to a smooth distribution for weights.




\section{Results}
\label{sec:Results}
As mentioned earlier, labeled data is not available at the isoform level. Based on what is done in the literature, first, the performance of trained models is computed at the gene level.

We compared the results of the proposed method with the state-of-the-art methods, including IMILP, WLRM, DeepIsoFun, DIFFUSE in Table \ref{tab:Evaluation_1}. This comparison is made on a subset of GO terms. Ninety-six GOs are left after pruning the GO terms with large size. Besides the first dataset (Dataset 1), two other datasets are used to have a comprehensive comparison. Dataset 2 consists of RNA-SEQ data for 29,806 human isoforms from 18,923 genes generated from 29 RSA studies from 455 experiments. Dataset 3 is composed of mouse isoforms, 17,191 isoforms corresponding to 13,962 genes, and 116 RSA studies in 365 experiments.

The results show that the DIFFUSE method is the best-proposed method by a few percent. However, it is a complex and high-dimensional method compared to the proposed method. Additionally, as it trains a separate model for each GO term, It requires more time and computational resources to train and inference.

\begin{center}
\begin{table}
\begin{tabular} { | c | c | c | c | c | c | c |}
\hline
Method & \multicolumn{2}{|c|}{Dataset 1} & \multicolumn{2}{|c|}{Dataset 2} & \multicolumn{2}{|c|}{Dataset 3} \\
\cline{2-7}
& AUC  & AUPRC  & AUC  & AUPRC & AUC  & AUPRC  \\
\hline
Our Method & 0.812 & 0.515 & 0.801 & 0.509 & 0.798 & 0.499\\
DIFFUSE & 0.835 & 0.585 & 0.828 & 0.537 & 0.817 & 0.524\\
DeepIsoFun & 0.729 & 0.280 & 0.722 & 0.257 & 0.712 & 0.231 \\
WLRM & 0.685 & 0.265 & 0.667 & 0.237 & 0.672 & 0.201\\
IMILP & 0.678 & 0.317 & 0.662 & 0.292 & 0.639 & 0.288\\
\hline
\end{tabular}
\caption{\label{tab:Evaluation_1}Comparing the proposed method with other isoform function prediction methods. IMILP performs classification in three classes instead of two; the third class is devoted to the unknown. we measure its AUC and AUPRC values using only the positive and negative classes. Our method's performance is comparable with the state-of-the-art method, although the computational complexity is much fewer.}
\end{table}
\end{center}

To find the impact of components of the proposed model in its current performance, we removed a few innovative components and reran the training and evaluation process. More specifically, we removed the customized max-pool layer over the child go term score, which injects the structure of go terms in the model. The average AUC and AUPRC dropped 1.4\% and 7.2\%, respectively. Moreover, we replaced the Protvec component with a dynamic embedding used in the literature (e.g., DIFFUSE). The average AUC and AUPRC dropped 5.4\% and 20.2\%, respectively.

The model's performance is evaluated at the isoform level in the next step. Functions for 14 isoforms of 6 genes have been determined. Out of the 14 isoforms, our method predicted correct functions for 10 of them. The comparison between our methods and literature is presented in the Table \ref{tab:Evaluation_2}.

\begin{table}
\centering
\begin{scriptsize}
\begin{tabular} {| m{1.2cm} || c | c | m{1.5cm} | m{1cm}  m{1cm}  m{1cm}  m{1cm}  c |}
    \hline
    \textbf{GO term} & \textbf{Gene} & \textbf{Isoform} & \textbf{Literature evidence} & \multicolumn{5}{c|}{\textbf{Prediction method}}\\
    \hline
    & & & & Our method & DIFFUSE & DeepIsoFun & WLRM & iMILP\\
     
    \textbf{GO: 0046872} & ACE & P12821-1 & + & + & + & + & + & +\\
    & & P12821-3 & + & + & + & - & - & +\\
    & ACMSD & Q8TDX5-1 & + & + & + & + & + & +\\
    & & Q8TDX5-2 & - & + & + & + & + & +\\
    & GCH1 & P30793-1 & + & + & + & + & - & +\\
    & & P30793-2 & - & - & - & - & - & -\\
    & & P30793-4 & - & - & + & + & + & +\\
    \textbf{GO: 0005634} & ADK & P55263-1 & + & + & + & + & + & -\\
    & & P55263-2 & - & + & + & - & + & -\\
    & AIFM1 & O95831-1 & + & + & + & + & + & +\\
    & & O95831-3 & - & + & - & + & + & -\\
    & & O95831-4 & - & - & - & - & - & +\\
    & PPP1R8 & Q12972-1 & + & + & + & + & + & +\\
    & PPP1R8 & Q12972-3 & - & + & - & - & - & +\\
    \hline
    \textbf{Accuarcy} & \multicolumn{3}{c|}{} & 71.4\% & 78.6\% & 71.4\% & 50.0\% & 64.3\% \\
    \hline
\end{tabular}
\end{scriptsize}
\caption{\label{tab:Evaluation_2}Performance of the proposed method and conventional methods on the known isoforms. Experimental evidence for
six genes have been found in the literature: ACE \cite{corradi2006crystal}, ACMSD \cite{pucci2007tissue}, GCH1 \cite{auerbach2000zinc}, ADK \cite{cui2009subcellular}, AIFM1 \cite{delettre2006identification} and PPP1R8 \cite{chang1999alternative}. Our model misclassifies four isoforms which are one sample more than the state-of-the-art.}
\end{table}


\section{Conclusion}
\label{sec:Conclusion}
This research proposed a deep neural network model to predict the isoform functions by their sequence, conserved domains, and expression profiles. Using a comprehensive model to train all GO terms in the same network led to a significant decay in time and computational resource usage. Moreover, the accuracy is close to the state-of-the-art models. However, the model's performance can be improved in several aspects. First, the expression profile data can be modeled as a graph in GCNs, and the inter-isoform relations can be presented better. Second, the results can be improved by applying other approaches and improvements in the Multi-Task Learning area. One way is to increase the task-specific layers. In our proposed method, the only task-specific layer is the final dense layer that computes the isoforms' score. The second way is to let the model choose the shared and task-specific layers in training. The last way is to use the methods proposed to cluster classes in the first place and then train a separate network for each cluster, which is a balance between training each class separately and training all of them in the same network.


\bibliography{bibs/full,bibs/refs}


\end{document}